\documentclass[aps,twocolumn,showpacs,preprintnumbers,amsmath,amssymb, A4]{revtex4}

\usepackage{color}

\usepackage{colordvi}
\usepackage{amssymb}
\usepackage{amsmath}
\usepackage{epsf}
\usepackage{mathrsfs}
\usepackage{graphicx}
\usepackage{appendix}

\def\nn{\nonumber}
\def\be{\begin{equation}}
\def\ee{\end{equation}}
\def\bea{\begin{eqnarray}}
\def\eea{\end{eqnarray}}
\begin{document}

\title{\textbf{Extended Bose Hubbard model for two leg ladder systems in artificial magnetic fields }}
\author{Rashi Sachdeva}
\email{rashi.sachdeva@oist.jp}
\affiliation{Quantum Systems Unit, Okinawa Institute of Science and Technology Graduate University, Okinawa 904-0495, Japan}
\author{Manpreet Singh \footnote{Present address: Department of Physics, Indian Institute of Technology, Guwahati-781039, Assam, India }}

\affiliation{Theoretical Physics Division, Physical Research Laboratory, Navrangpura, Ahmedabad 380009, India}

\author{Thomas Busch}
\affiliation{Quantum Systems Unit, Okinawa Institute of Science and Technology Graduate University, Okinawa 904-0495, Japan}

\date{\today}

\begin{abstract}
We investigate the ground state properties of ultracold atoms with long range interactions trapped in a two leg ladder configuration in the presence of an artificial magnetic field. Using a Gross-Pitaevskii approach and a mean field Gutzwiller variational method, we explore both the weakly interacting and strongly interacting regime, respectively. We calculate the boundaries between the density-wave/supersolid and the Mott-insulator/superfluid  phases as a function of magnetic flux and uncover regions of supersolidity. The mean-field results are confirmed by numerical simulations using a cluster mean field approach. 
 \end{abstract}

\maketitle
  
\section{Introduction}
Ultracold atoms in optical lattices are a topic of significant interest, as they can realize the fundamental models of periodic many body physics which are difficult to explore in traditional condensed matter systems. Typical experiments are highly tunable with respect to the lattice strength and the inter-particle interactions and therefore  allow to explore a wide parameter range~\cite{lewenstein_book}.
This has opened up many new avenues for the study of quantum phase transitions, in particular when combined with
the recent progress in creating  artificial magnetic fields for ultracold atoms in discrete~\cite{magfield_OL} as well as continuum systems~\cite{magfield_cont}. Such systems have been studied using the Bose Hubbard model~\cite{jaksch, bloch} with a complex tunneling term, which incorporates the effect of artificial magnetic fields for bosonic gases in optical lattices. The main effect of these magnetic fields can be observed even without interactions and the single particle spectrum for bosons in a periodic potential in presence of strong magnetic field forms a self-similar structure known as the Hofstadter butterfly~\cite{hofstadter}.

In addition, it has been proposed that a system of polar gases in optical lattices can give rise to a crystalline phase due to the long
range van der Waal type interaction~\cite{pfaureview}, and certain conditions make it possible to stabilize the
exotic supersolid phase~\cite{dorneich,prokofev,luthra}. The observation of Bose Einstein condensation in chromium
~\cite{Griesmaier} followed by
the realization of degenrate quantum gases made from other highly-magnetic
species, including dysprosium~\cite{Lev}
and erbium~\cite{Aikawa},  and experiments on polar molecules, such as KRb~\cite{KKNi}, have therefore opened up possibilities for manipulating the
off-site interactions in optical lattices.

Another aspect of quantum gases is the possibility to create low dimensional systems, where $1D$ or quasi-$1D$ systems are of special interest as
they can enhance the role interactions play, which is crucial for realizing novel phases~\cite{rigol_giamarchi_rmp}.  
 Among these, quasi-$1D$  ladders systems hold particular importance because the extra coupling between the legs of the ladder 
introduces an additional degree of freedom that can significantly influence quantum phase transitions even in a simple model like the Bose-Hubbard ladder~\cite{giamarchi1}.
 In the presence of small magnetic fields, the two leg bosonic ladder model shows an analog of the Meissner state, where currents circulate along the legs of the ladder, while for higher magnetic fields, currents also start flowing along the rungs, resulting in a vortex state. This phase transition from Meissner to vortex state has recently been experimentally observed by M.~Atala et al.~\cite{bloch_ladder}. Related theoretical works~\cite{giamarchi_meinereffect, oktel_ladder} have studied other non-trivial effects of artificial magnetic fields on such system, including situations where the constraints of the complicated single particle spectrum or the rationality of the applied magnetic field are absent. 

This recent progress in the experimental realization of artificial magnetic fields, long range interacting Bose-Einstein condensates (BECs), and low dimensional ladder systems, allows to ask interesting question about the interplay of these three influences on cold atomic systems. To investigate this, we study the specific example of long range interacting BEC systems on a {\it two} leg ladder in the presence of a {\it uniform} artificial magnetic field and find that long range interactions gives rise to two phases, namely a density-wave and a supersolid phase~\cite{rashi1,rashi2}, in addition to the Mott-insulator and superfluid phases obtained in usual Bose Hubbard model~\cite{jaksch}. 
For weak interactions, we find that the transition from the Meissner to the vortex phase moves to higher magnetic field values, whereas for strong interactions the system displays density-wave and Mott-insulator phases. Both of these are stabilized by the magnetic fields~\cite{rashi1,rashi2}. In addition, our numerical calculations also show the appearance of a supersolid phase outside the density-wave lobes, which stems from the competition between the short and the long range repulsive interactions. 
 
The manuscript is organized as follows. In Section \ref{Sec: ebhm} we introduce the extended Bose Hubbard model (eBHM) with a two-leg ladder geometry in an artificial magnetic field and in Section \ref{Sec: singleparticle} we review the properties of its single particle spectrum. The weakly interacting regime is then discussed using a Gross-Pitaevskii approach  in Section \ref{Sec: GPapproach}, where we also study the excitations of the system beyond the mean field calculations. In section \ref{Sec: Gutzwiller}, we present the analytic calculations for the determination of the phase boundaries using the variational Gutzwiller approach and Section \ref{Sec: ClusterMF} presents the numerical calculations performed using the cluster mean field theory. Finally, in Section \ref{Sec: Summary} we present the summary and outlook of the work done. 

\section{ Extended Bose Hubbard model for two-leg ladder}
\label{Sec: ebhm}

\begin{figure}[!htbp]
   \centering
\includegraphics[width=.5\textwidth]{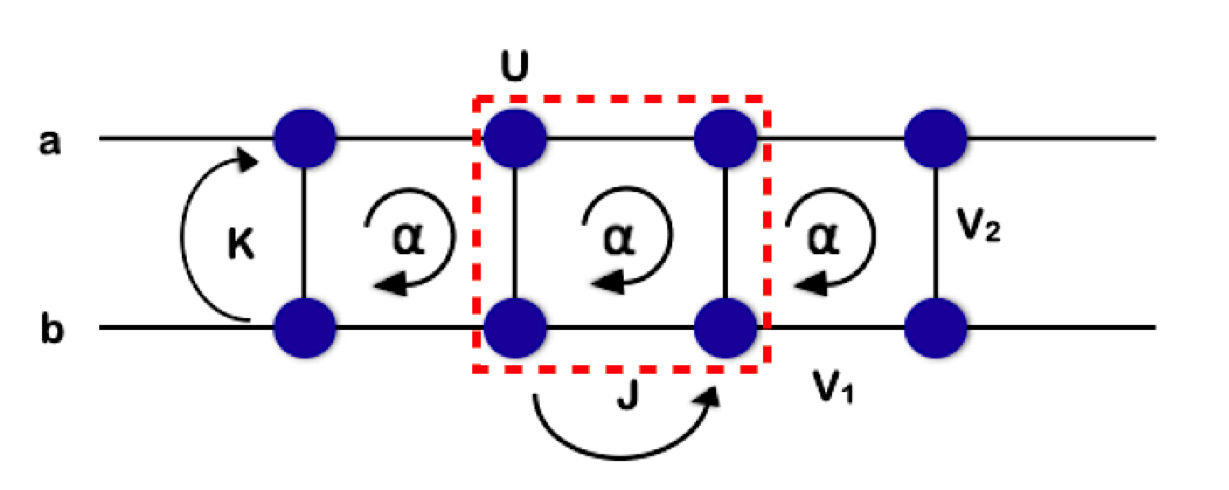}
\caption{(Color online) Schematic of the two-leg ladder system with long range (nearest neighbour) interactions in the presence of a uniform magnetic flux. The dashed red box indicates a single unit cell for the cluster mean field calculations described in Section \ref{Sec: ClusterMF}.}
\label{fig:fig_ladder}
\end{figure}

The Hamiltonian describing bosons in a two-leg ladder geometry in the presence of magnetic flux can be written as
\begin{align}
  H=&-J\sum_{i}\left[(e^{-i\alpha}a_i^\dagger a_{i+1}+e^{i\alpha}b_i^\dagger b_{i+1})+ h.c.\right]\nonumber\\
  &- K\sum_{i}(a_i^\dagger b_i+ h.c.)+{U \over 2} \sum_{i,p} n_i^p(n_i^p-1)\nonumber\\
  &+V_1 \sum_{i,p} n_i^p n_{i+1}^p+V_2 \sum_{i}n_i^a n_i^b-\mu\sum_{i,p}n_i^p\;,
  \label{eq:eq1_model}
\end{align}
where the $p_i (p_i^\dagger)$ are the bosonic annihilation (creation) operators at site $i$ of leg $p~(=a,b)$,
$n_i^p$ is the number operator at site $i$ of leg $p$, $\alpha$ is the magnetic flux and $\mu$ is the chemical potential. 
The intra- and inter-leg hopping amplitudes are given by $J$ and $K$,  the on-site interaction energy between two atoms is given by $U$ and $V_1$ and $V_2$ 
are non-local nearest-neighbour ineractions (NNI), along a leg and between the legs respectively.
The ratios $J/U$ and $K/U$ can be changed in an experiment by tuning the laser intensities and/or varying the separation between the legs (see Fig.~\ref{fig:fig_ladder}).
Similarly, $V_1$ and $V_2$ can be tuned by changing the polarizing angle of the external fields relative to the plane of the ladder. We assume up down symmetry for the ladder, which implies that the chemical potential $\mu$ and the onsite interactions $U$ are identical for each leg of ladder. The phase $\alpha$ in the hopping terms is given by $\alpha=(e/\hbar)\int_{r_i}^{r_j}d\mathbf{r}\cdot\mathbf{A(r)}$, where $\mathbf{A(r)}$ is the vector potential giving rise to the magnetic field $\mathbf{B}=\nabla \times\mathbf{A}$. We use the Landau gauge $\mathbf{A(r)}=-By\hat{x}$ and write the phase $\alpha$ in the form $\alpha=\pi\phi/\phi_0$, where $\phi$ is the magnetic flux through each plaquette, with $\phi_0=h/e$ as the flux quantum.

The two leg ladder system is quite advantageous compared to fully two dimensional systems, in the sense that it can assume any value between $0$ and $1$ for the magnetic flux $\alpha/\pi$, whereas in two-dimensional systems $\alpha/\pi $ has to be a rational number. This is because of symmetry breaking due to the specific gauge choice, which can be restored in a $q$-fold enlarged magnetic unit cell. Hence, without the constraint of rational magnetic field values, the two leg ladder system is less complex and can be used to study non-trivial effects of magnetic fields in a relatively simple way. In fact, the effect of the presence of magnetic fields can already be studied at the non-interacting single particle level, which is discussed in next section.

\section{Single particle spectrum}\label{Sec: singleparticle}
In this section, we briefly review the structure of the single particle energy spectrum as a function of the magnetic flux values. The bottom of the lowest band is known to show two types of topologies: a single minimum for small flux values and two minima for large flux values.

\begin{figure*}[!htbp]
   \centering
\includegraphics[width=1\textwidth]{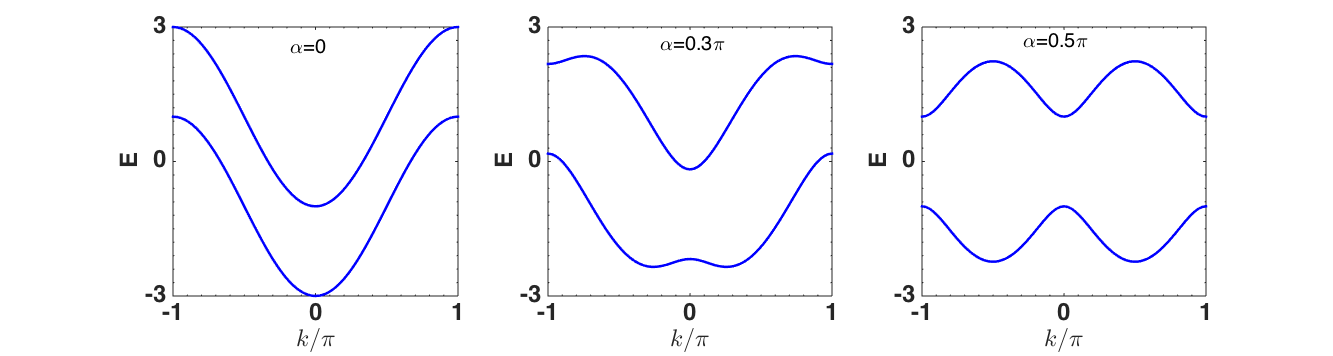}
\caption{(Color online) Single-particle spectrum of the two-leg ladder system for different magnetic flux strengths $\alpha$, and fixed interleg-to-intraleg hopping ratio $K/J=1$. The two lowest bands are shown.}
\label{Fig: singleparticle_dispersion}
\end{figure*}

To see this, we assume $U=V_1=V_2=0$ and write the Fourier components of the field operators $a_j$ and $b_j$  as
\be a_j=\frac{1}{\sqrt{L}}\sum_{k}a_ke^{ikj}, ~b_j=\frac{1}{\sqrt{L}}\sum_{k}b_ke^{ikj}, \label{fourier}\ee
where $[a_k,a_{k'}^{\dag}]=\delta_{k,k'}$ and $[b_k,b_{k'}^{\dag}]=\delta_{k,k'}$. Using this transformation, the Hamiltonian in Eq.~\eqref{eq:eq1_model} takes the form
\be H=-\sum_{k}[\xi_{ak}a_{k}^{\dag}a_k+\xi_{bk}b_{k}^{\dag}b_k +K(a_k^{\dag}b_k+b_k^{\dag}a_k)], \ee
where $\xi_{ak}=2J\text{cos}(k-\alpha)$ and $\xi_{bk}=2J\text{cos}(k+\alpha)$. To diagonalize the Hamiltonian, we perform a Bogoliubov transformation $A_k=\text{cos}\theta~a_k+\text{sin}\theta~b_k$, $B_k=-\text{sin}\theta~ a_k+\text{cos}\theta~ b_k$ with $\theta=\frac{1}{2} \text{tan}^{-1}\bigg(\frac{2K}{\xi_{ak}-\xi_{bk}}\bigg)$, which allows to find the energy eigenvalues as 
\be \epsilon_{1,2}=-2~\text{cos}k~\text{cos}\alpha\mp\sqrt{\tilde{K}^2+4~\text{sin}^2 k~\text{sin}^2\alpha}, \ee
where $\tilde{K}=K/J$, and we normalize the energy to the interleg hopping $J$.  The above expression for eigenvalues gives a two-band structure as well as the $2\pi$ periodicity, reflecting the two-leg ladder geometry.  It is shown for different  magnetic flux values in Fig.~\ref{Fig: singleparticle_dispersion}, and one can see that with increasing field strength the single  band minimum  at $k=0$ evolves  into two at non-zero $k$ values, which are degenerate and symmetric about the origin. The critical value of the magnetic field for the split is given by
\be \alpha_{c}=\text{cos}^{-1}~\left(-\frac{\tilde{K}}{4}\pm\sqrt{\frac{\tilde{K}^2}{16}+1}\right).\ee
Above this value, the ground state of the system is no longer spatially uniform, but will be a superposition of plane waves corresponding to the two minima of the dispersion relation. In the experiment by M.~Atala et al.~\cite{bloch_ladder}, these two ground states were observed for weakly interacting bosons and termed as Meissner and vortex phases.

\begin{figure}[tb]
\includegraphics[width=0.9\columnwidth]{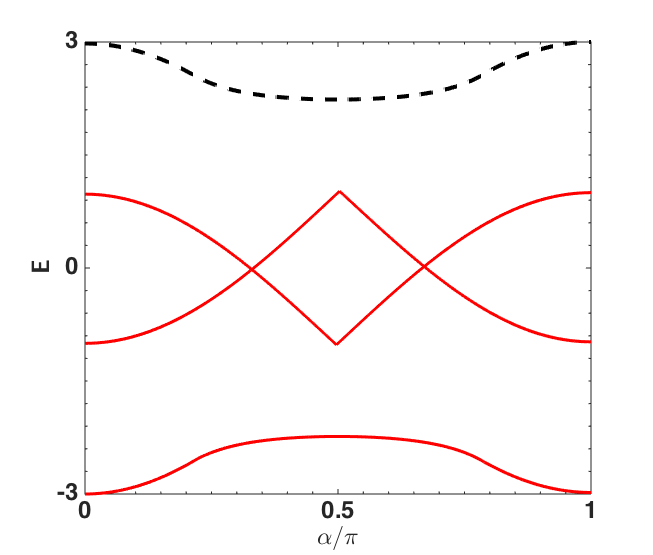}
\caption{(Color online) Band diagram (\textit{Hofstadter's butterfly}) for the non-interacting two leg ladder bosonic system.}
\label{fig:hofs}
\end{figure}

To recover the \textit{Hofstadter butterfly} for this two-leg ladder system we plot the minimal and maximal value of each band for each value of magnetic flux in Fig.~\ref{fig:hofs}. The spectrum can be seen to be symmetric around $\alpha=0.5\pi$ and the highest eigenvalue, which is of special importance for the analytical calculations presented below, is  emphasised by the dashed black line.

\section{Gross-Pitaevskii approach for weakly interacting system}\label{Sec: GPapproach}

While our ultimate aim is to investigate the effect of strong interactions on the transition between different phases, we will in the following first
explore the effects of weak interactions on the system. 
When the interactions and magnetic field strengths are small, the system stays in the superfluid state and the hopping term is dominant in the Hamiltonian. Thus, assuming that the fluctuations in the condensate are negligible, we can make the usual mean-field approximation
\bea a_i\rightarrow \langle a_i\rangle&=&\psi_i, \nn\\
b_i\rightarrow \langle b_i\rangle&=& \phi_i, \label{fluc}\eea
which leads to the following functional form of the energy ($J=1$)
\begin{widetext}
\bea E\left[\{\psi_j\},\{\phi_j\}\right]&=& -\sum_{j}\big[e^{-i\alpha}\psi_j^{\ast}\psi_{j+1}+e^{i\alpha}\phi_j^{\ast}\phi_{j+1}+K\psi_{j}^{\ast}\phi_j+\text{c.c}\big]+\frac{U}{2}\sum_j\big[\psi_j^{\ast}\psi_j(\psi_j^{\ast}\psi_j-1)+\phi_j^{\ast}\phi_j(\phi_j^{\ast}\phi_j-1)\big]\nn\\
& & +\sum_j\big[V_1\psi_j^{\ast}\psi_j\psi_{j+1}^{\ast}\psi_{j+1}+V_2\psi_j^{\ast}\psi_j\phi_j^{\ast}\phi_j+V_1\phi_j^{\ast}\phi_j\phi_{j+1}^{\ast}\phi_{j+1}\big]-\mu\sum_j\left[\psi_j^{\ast}\psi_j+\phi_j^{\ast}\phi_j\right].\label{enfunc}
\eea
The variation of the energy functional $i\partial\psi_i/\partial t=\delta E/\delta\psi_i^{\ast}$ and $i\partial\phi_i/\partial t=\delta E/\delta\phi_i^{\ast}$ around the minimum then gives the coupled Gross Pitaevskii equations for the order parameter $\Psi=( \psi_j ,\phi_j)^{T}$ as
\bea i\frac{\partial \psi_j}{\partial t}&=& -\big[e^{-i\alpha}\psi_{j+1}+e^{i\alpha}\psi_{j-1}+K\phi_j\big]+U|\psi_{j}|^{2}\psi_j-\left(\frac{U}{2}+\mu\right)\psi_j+V_{1}|\psi_{j+1}|^{2}\psi_j+V_2|\phi_j|^{2}\psi_j,\nn\\
i\frac{\partial \phi_j}{\partial t}&=& -\big[e^{i\alpha}\phi_{j+1}+e^{-i\alpha}\phi_{j-1}+K\psi_j\big]+U|\phi_{j}|^{2}\phi_j-\left(\frac{U}{2}+\mu\right)\phi_j+V_{1}|\phi_{j+1}|^{2}\phi_j+V_2|\psi_j|^{2}\phi_j.\nn\\
\label{coup_gpe}\eea
\end{widetext}
The chemical potential can be straightforwardly determined from the zeroth order terms as 
\bea \mu =-(2~\text{cos}\alpha+K)+0.5U(2n-1)+(V_1+V_2)n, \label{chempot}\eea
and the excitation spectrum can be found by taking higher order fluctuations into account and writing the order parameter as 
$\Psi=\Psi_0+\delta\Psi$, which takes the  form

\bea \psi_j&=&\sqrt{n}+Ae^{i(kx_j-\omega t)}+B^{\ast}e^{-i(kx_j-\omega t)}, \nn\\
\phi_j&=&\sqrt{n}+Ce^{i(kx_j-\omega t)}+D^{\ast}e^{-i(kx_j-\omega t)}. \label{fluc_exp} \eea
Here $x_j$ is the position of the lattice site $j$, $k$ is the reciprocal lattice vector and $A$, $B$, $C$ and $D$ are the small complex parameters. 

\begin{figure*}[tb]
\includegraphics[width=15cm]{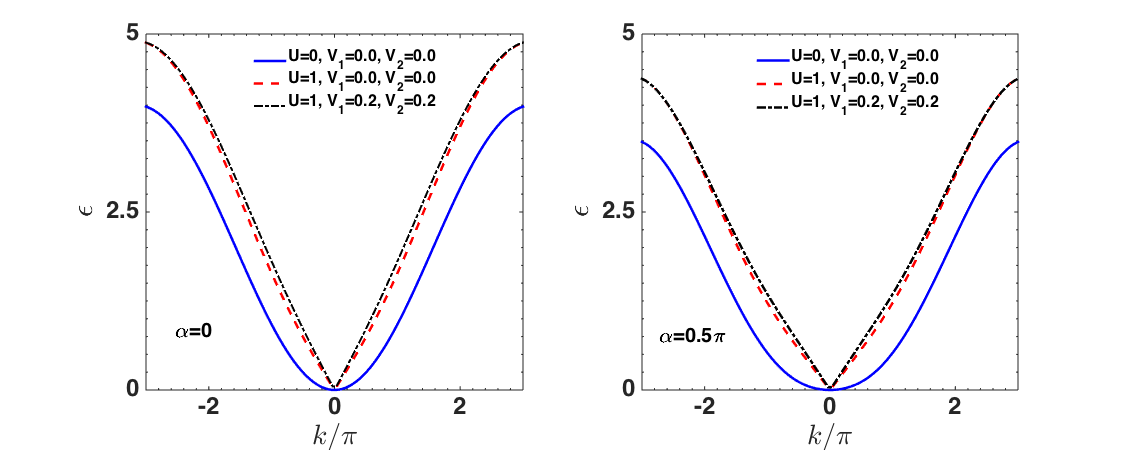}%
\caption{(Color online) Band diagram calculated within the Gross Pitaevskii approximation for $\alpha=0$ (left) and $\alpha=0.5\pi$ (right).}
\label{band_spec}
\end{figure*}

Substitituting these expressions into the coupled Gross-Pitaevskii equations \eqref{coup_gpe}, we get an algebraic equation of the form $H_{g}\delta\Psi=\omega\delta\Psi$ with $\delta\Psi=(A, B, C, D)^{T}$, where the roots $\omega$ can be calculated by solving the determinant
\begin{widetext}
\[ \begin{vmatrix}
-\xi_{ak}-\omega&Un+V_1ne^{ika} & -K+V_2n & V_2n \\
-Un-V_1ne^{ika} &\xi_{bk}-\omega & -V_2n & K-V_2n \\
-K+V_2n & V_2n & -\xi_{bk}-\omega & Un+V_1ne^{ika} \\
-V_2n & K-V_2n & -Un-V_1n e^{ika} & -\xi_{ak}-\omega
\end{vmatrix}
=0\]
\end{widetext}
Here $\xi_{ak}=[2~\text{cos} (ka-\alpha)-2~\text{cos}\alpha-Un-K-V_1ne^{ika}]$ and $\xi_{bk}=[2~\text{cos} (ka+\alpha)-2~\text{cos}\alpha-Un-K-V_1ne^{ika}]$. This gives the spectrum as a function of different values for the flux $ \alpha$ and the repulsive interactions $ U$, $V_1$ and $V_2$. The resulting bands  for the situation with no interactions, only onsite interactions and both onsite and nearest neighbour repulsive interactions in the system are shown in Figure \ref{band_spec}.  One can immediately see that for no magnetic flux the repulsive interactions lead to a phonon-like spectrum around $k=0$, however  this effect is weakened for finite magnetic flux values. Hence the magnetic flux and the replusive interactions compete with each other in deciding the shape of the spectrum. Since the Hamiltonian in Eq.~\eqref{eq:eq1_model} also possesses long range repulsive interactions in addition to the onsite interactions, it can be expected that higher magnetic fields are needed to counter the effects of the interactions compared to the situation where only onsite interactions are present \cite{oktel}.  

\section{Variational Mean field approach : strong interaction limit} \label{Sec: Gutzwiller}
\subsection{Mott-Insulator/Superfluid phase boundaries}

\begin{figure}
\center
\includegraphics[width=0.5\textwidth]{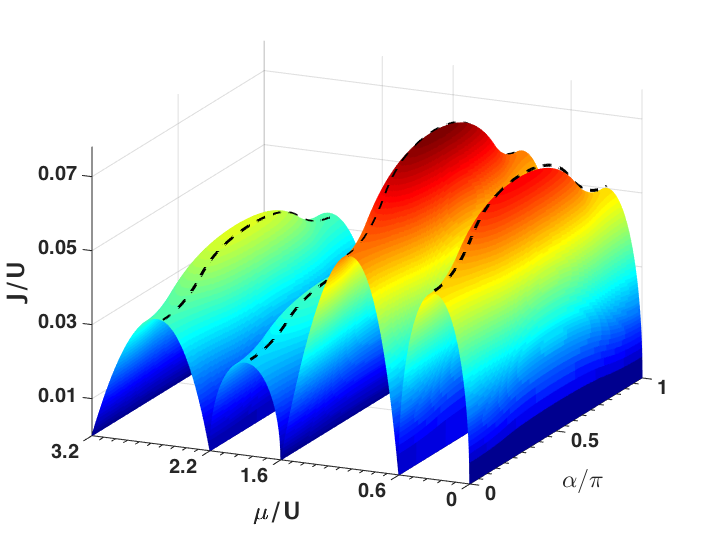}
\caption{(Color online) Full phase diagram for the eBHM for a two leg ladder ($J/U$ vs. $\mu/U$) as a function of magnetic flux $\alpha$. The dashed black lines inversely follow the form of the highest eigenvalue of the non-interacting band spectrum as shown in Fig~\ref{fig:hofs}.}
\label{fig:V_R}
\end{figure}

In the following we will explore the transition from the Mott-insulator (MI) to the superfluid (SF) regime as a function of $J$, $K$, $V_1$, $V_2$, $\mu$ and $\alpha$. Here we scale the Hamiltonian in Eq.~\eqref{eq:eq1_model}  by setting $U=1$. For the perfect Mott-insulating phase, the wavefunction is localized with an equal number of particles, $n_0$, at each site. The phase boundary of this phase can then be 
analytically determined by calculating the energy for particle-hole-type excitations using a reduced-basis variational ansatz for the Gutzwiller wave function. This implies that we work very near to the phase boundary, i.e in the strongly interacting regime. Very close to the MI phase boundary only Fock states close to the MI one are populated and we can write the Gutzwiller ansatz for the local sites as
\begin{align}
	|G\rangle_{al}  &= f_{n_0-1}^{al}|n_0-1\rangle+f_{n_0}^{al}|n_0\rangle+f_{n_0+1}^{al}|n_0+1\rangle, \\
	|G\rangle_{bl}  &= f_{n_0-1}^{bl}|n_0-1\rangle+f_{n_0}^{bl}|n_0\rangle+f_{n_0+1}^{bl}|n_0+1\rangle.  
\end{align}
The total wave function of the system is then given by the product state $|\Psi\rangle= \Pi_l |G\rangle_{al}  |G\rangle_{bl}  $, where $|G\rangle_{al}$ and $|G\rangle_{bl} $ are the wave functions of each rung.
Furthermore we parametrize

\begin{align} 
(f_{n_0-1}^{al}, f_{n_0}^{al}, f_{n_0+1}^{al})&=(e^{-i\theta_{l}}\Delta_{al},\sqrt{1-\Delta_{al}^2-\Delta_{al} ^{'2}},e^{i\theta_{l}}\Delta_{al}^{'}) \nn\\
(f_{n_0-1}^{bl}, f_{n_0}^{bl}, f_{n_0+1}^{bl})&=(e^{-i\theta_{l}}\Delta_{bl},\sqrt{1-\Delta_{bl}^2-\Delta_{bl} ^{'2}},e^{-i\theta_{l}}\Delta_{bl}^{'})\nn\\
\label{coefficients}
\end{align}
with complex variational parameters $\Delta_{al}, \Delta_{al} ^{'},\Delta_{bl}, \Delta_{bl}^{'}\ll1$ to ensure the normalisation condition of states $ |G_{al} \rangle$ and $|G_{bl} \rangle$. Minimizing the energy functional with respect to the variational parameters $\Delta_{a,l}, \Delta_{a,l} ^{'},\Delta_{b,l}, \Delta_{b,l}^{'}$ and $\theta_{l}$  (see Appendix \ref{misf_app} for details), then gives the following simple relation for the boundary of the Mott-insulator phase
\begin{widetext}
\be J_c=\frac{(n_0-\mu)(1-n_0+\mu)+n_0(2V_1+V_2)(1-2n_0+2\mu)-n_0^2(2V_1+V_2)^2}{\lambda_F(\mu+1)-n_0 \lambda_F(2V_1+V_2)} \label{tcMi}, \ee
\end{widetext}
for any value of the chemical potential $\mu$, for given strengths $V_1$ and $V_2$ of the NNI, and for any value of the magnetic flux through the ladder. The dependence on magnetic flux is implicit in the largest eigenvalue, $\lambda_{F}$, of the single particle Hamiltonian $\textbf{F}$ and the Mott-insulator/superfluid phase boundaries are shown as a function of the magnetic flux $\alpha/\pi$ as the second and fourth lobe in Fig.~\ref{fig:V_R}. 

It is clear from Eq.~\eqref{tcMi} that the highest eigenvalue of the band diagram for the non-interacting system (dashed black line in Fig.~\ref{fig:hofs}) inversely relates to the critical transition boundaries (dashed black line in Fig.~\ref{fig:V_R}). In the next section we determine the density-wave/supersolid phase boundaries using a similar variational approach. 

\subsection{Density-wave/Supersolid phase boundaries}

The phase boundary of the density-wave (DW) to supersolid (SS) phase can again be determined analytically by calculating the energy of the particle-hole-type
excitations using a reduced-basis variational ansatz for the Gutzwiller wave function near the phase boundary.  The density-wave phase is characterized by an alternate number density distribution where neighboring lattice sites are occupied by $n_0$ and $m_0$ particles in a periodic way. It can therefore be viewed as consisting of two sublattices, $A$ and $B$, with $n_0$ and $m_0$ particles on each site of the respective sublattices. Hence we can write the total wave function for the system as the product of wavefunctions for sublattice $A$ and sublattice $B$. For the two leg ladder system, where we already label the legs by $a$ and $b$, this requires to split each ladder further into a sublattice structure, leading to a total wavefunction that is the product of four individual wavefunctions,  $|\Psi\rangle=\Pi_l |G \rangle_{a_1l}  |G\rangle_{a_2l} |G \rangle_{b_1l}  |G\rangle_{b_2l} $.

Since we are only interested in the region very close to the DW phase boundary, we again limit the ansatz for the wavefunctions to Fock states that have at most one particle added or removed from the equilibrium number and choose
\begin{align}
	 |G\rangle_{a_1l}  &= f_{n_0-1}^{a_1l}|n_0-1\rangle+f_{n_0}^{a_1l}|n_0\rangle+f_{n_0+1}^{a_1l}|n_0+1\rangle \nn\\
	|G\rangle_{a_2l}  &= f_{m_0-1}^{a_2l}|m_0-1\rangle+f_{m_0}^{a_2l}|m_0\rangle+f_{m_0+1}^{a_2l}|m_0+1\rangle\nn\\  
	|G\rangle_{b_1l}  &= f_{m_0-1}^{b_1l}|m_0-1\rangle+f_{m_0}^{b_1l}|m_0\rangle+f_{m_0+1}^{b_1l}|m_0+1\rangle \nn\\
	|G\rangle_{b_2l}  &= f_{n_0-1}^{b_2l}|n_0-1\rangle+f_{n_0}^{b_2l}|n_0\rangle+f_{n_0+1}^{b_2l}|n_0+1\rangle. 
\end{align}

\begin{figure}
\center
\includegraphics[width=.5\textwidth]{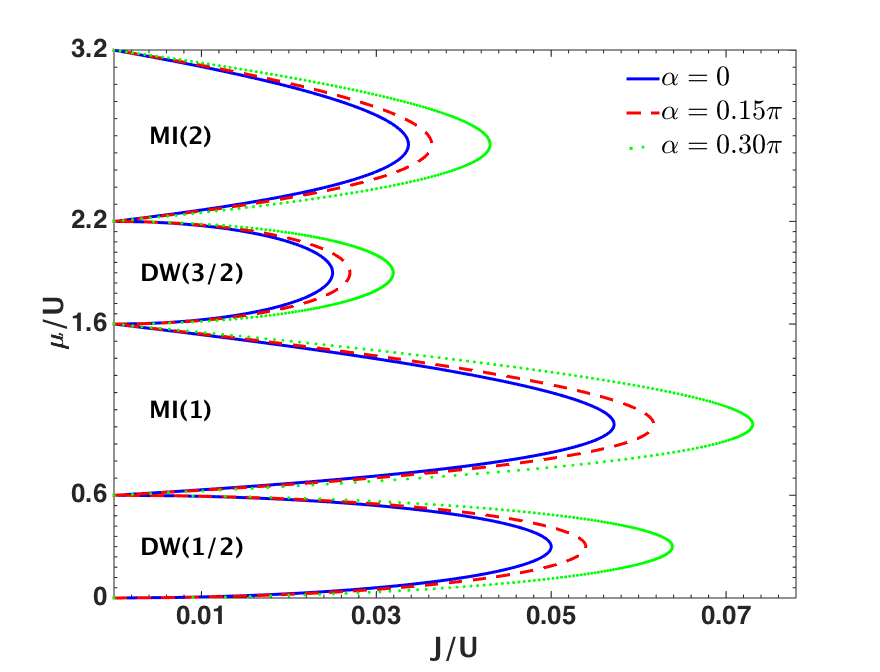}
\caption{(Color online) Phase diagram of the eBHM for the two leg ladder for different values of magnetic flux $\alpha$, for $K=1$, $U=1$ and $V_1=V_2=0.2$, calculated using a variational mean field approach. The DW and MI phases are indicated with their average occupancy per site in brackets.}
\label{fig_ebhmfull}
\end{figure}

Following a similar procedure as in the previous section, we parameterize 
\begin{widetext}
\begin{align} 
(f_{n_0-1}^{a_{1}l}, f_{n_0}^{a_{1}l}, f_{n_0+1}^{a_{1}l})&=(e^{-i\theta_{l}}\Delta_{a_{1}l},\sqrt{1-\Delta_{a_{1}l}^2-\Delta_{a_{1}l} ^{'2}},e^{i\theta_{l}}\Delta_{a_{1}l}^{'}),\\
 (f_{m_0-1}^{a_{2}l}, f_{m_0}^{a_{2}l}, f_{m_0+1}^{a_{2}l})&=(e^{-i\theta_{l}}\Delta_{a_{2}l},\sqrt{1-\Delta_{a_{2}l}^2-\Delta_{a_{2}l} ^{'2}},e^{i\theta_{l}}\Delta_{a_{2}l}^{'}),\\
 (f_{m_0-1}^{b_{1}l}, f_{m_0}^{b_{1}l}, f_{m_0+1}^{b_{1}l})&=(e^{-i\theta_{l}}\Delta_{b_{1}l},\sqrt{1-\Delta_{b_{1}l}^2-\Delta_{b_{1}l} ^{'2}},e^{i\theta_{l}}\Delta_{b_{1}l}^{'}),\\
 (f_{n_0-1}^{b_{2}l}, f_{n_0}^{b_{2}l}, f_{n_0+1}^{b_{2}l})&=(e^{-i\theta_{l}}\Delta_{b_{2}l},\sqrt{1-\Delta_{b_{2}l}^2-\Delta_{b_{2}l} ^{'2}},e^{i\theta_{l}}\Delta_{b_{2}l}^{'}), 
 \end{align}
 \end{widetext}
 with complex variational parameters $\Delta_{a_{1,2}l}, \Delta_{a_{1,2}l} ^{'},\Delta_{b_{1,2}l}, \Delta_{b_{1,2}l}^{'}\ll1$ to ensure the normalisation condition of the states $ |G_{a_{1}l} \rangle, |G_{a_{2}l} \rangle,|G_{b_{1}l} \rangle$ and $|G_{b_{2}l} \rangle$. Minimizing the energy functional with respect to the variational parameters is straightforward, though tedious (see Appendix \ref{dwss_app}), and allows to determine the boundary between the density-wave and the supersolid phase. 
  
The full phase diagram of extended Bose Hubbard model for different values of the magnetic flux $\alpha$ with $V_1=V_2=0.2$, $K=1$, and $U=1$ is shown in  Fig.~\ref{fig_ebhmfull}. One can see that a larger magnetic field increases the regions where the density-wave and the Mott-insulator phase appear by shifting the critical point $(J/U)_\text{crit}$ or tip of the lobe for the gapless-gapped phase transition to larger values. As in previous sub-section, we can again relate the highest eigenvalue of the band diagram with the critical transition boundaries of density-wave/supersolid phase (first and third lobe), which is again shown by the dashed black  line in Fig. \ref{fig:V_R}.
 
This enlargement of the insulating phases is expected since the effect of the magnetic field is to localize the single particle motion even for non-interacting systems, thus making the transition to an insulating phases easier.
Starting from a pure density-wave or Mott state, all excitations are mobile and move on a constant background of filled sites. The magnetic field tries to limit this mobility, which results in increasing size of the Mott lobes.   
It is worth noting that these results are exact within mean field theory. The shape of the DW and MI lobe is concave and independent of the dimensionality, since in the mean field calculations the dimensionality enters only through a prefactor. 
 However one cannot expect the mean field theory to be quantitatively accurate for quasi one-dimensional systems, since fluctuations are known to be particularly important in lower dimensions. Hence, the results from the above analysis carry only qualitative importance, and provide a general idea of how the phase boundaries are affected by the presence of magnetic flux. In particular, they can be expected to work only for small hopping strengths when correlations are small. We will therefore in the following present numerical calculations that allow for better accuracy.

\section{Numerical Results} \label{Sec: ClusterMF}
In this section we solve the eBHM for the two-legged ladder system using a self-consistent cluster mean-field theory (CMFT). This is
a method that has been used extensively to study quantum phase transitions in optical lattices and superlattices
~\cite{penna,macintosh,manpreet_ladder1} and
is known to allow for capturing the essential physics and obtaining full phase diagrams which
can match the results from more sophisticated and numerically exact methods in accuracy~\cite{Huerga,Yamamoto,Hassan}.
It considers a small cluster of sites which forms a unit cell for the complete system and treats the hopping terms within the cluster 
 as exact, while the bonds between adjacent clusters are assumed to be decoupled. In our approach we consider clusters including four lattice sites, see dashed red cell in Fig.~\ref{fig:fig_ladder}.

In this model the Hamiltonian given in Eq.~\eqref{eq:eq1_model} can be written as
\begin{equation}\label{eq:manp_eq1}
    H=\sum_r H^r_c,
\end{equation}
where $r$ is the summation index over all clusters and $H^r_c$ is the Hamiltonian for $r^{th}$ cluster, which can be 
written as (dropping the index $r$ since all the clusters are identical)
\begin{equation}
H_{c}=H_{e}+H_{d}.
\end{equation}
Here $H_e$ describes the exact part of the Hamiltonian and $H_d$ the decoupled part. Decoupling is performed at the sites which
connect to the adjacent clusters by using the mean-field decoupling approximation 
\begin{eqnarray}
{a_{i}^{\dagger}a_{j}}~\simeq~\langle{a_{i}^{\dagger}}\rangle{a_j}+
{a_i}^{\dagger}\langle{a_j}\rangle-\langle{a_i}^{\dagger}\rangle\langle{a_j}\rangle,\nonumber\\
{n_{i}^{\dagger}n_{j}}~\simeq~\langle{n_{i}^{\dagger}}\rangle{n_j}+
{n_i}^{\dagger}\langle{n_j}\rangle-\langle{n_i}^{\dagger}\rangle\langle{n_j}\rangle,
\label{eq:manp_eq2}
\end{eqnarray}
where $i$ and $j$ are adjacent sites in directly neighbouring clusters. 
Similar to Section \ref{Sec: GPapproach} the superfluid order parameter and occupation number density are defined as 
\begin{align}
\phi_{i}&\equiv\langle{a_{i}^{\dagger}}\rangle\equiv\langle{a_i}\rangle,\nonumber\\
\rho_{i}&\equiv\langle{n_{i}}\rangle,
\label{eq:manp_eq3}
\end{align}
so that the decoupled part of the Hamiltonian takes the form
\begin{eqnarray}
{{H}}_{d}=-J\sum_{p,i,i'}&\Big[&e^{-i\alpha}({\phi_i^p}^*p_{i'}+{\phi_{i'}^p} p_i^\dagger-\phi_{i'}^p {\phi_i^p}^*)\nonumber\\
&+&e^{i\alpha}({\phi_{i'}^p}^*p_{i}+\phi_i^p p_{i'}^\dagger-\phi_i^p {\phi_{i'}^p}^*)\Big]\nonumber\\
&+&V\sum_p (n_{i'}^p \rho_i^p-\rho_{i'}^p\rho_{i}^p),
\label{eq:manp_eq4}
\end{eqnarray}
where $p=\{a,b\}$ is the same as in section ~\ref{Sec: ebhm} and $\{i,i'\}=\{1,2\},~i\ne i'$.
The exact part ${{H}}_{e}$ of the Hamiltonian is given by 
\begin{eqnarray}
{{H}}_{e}=&-&J\Big[(e^{-i\alpha}a_1^\dagger a_{2}+e^{i\alpha}b_1^\dagger b_{2})+ h.c.\Big]\nonumber\\
&-& K\sum_{i=1}^2(a_i^\dagger b_i+ h.c.)+{U \over 2} \sum_{\substack{i=1\\p=a,b}}^2 n_i^p(n_i^p-1)\nonumber\\
&+&V_1 \sum_{p=a,b} n_1^p n_2^p+V_2 \sum_{i=1}^2 n_i^a n_i^b\nonumber\\
&-&\mu\sum_{\substack{i=1\\p=a,b}}^2 n_i^p.
\label{eq:manp_eq5}
\end{eqnarray}
We again set the energy scale by choosing $U=1$ and work in the occupation number basis.  The Hamiltonian matrix can then be constructed using the expression given
by $H_{c}$ and diagonalized self-consistently to obtain the ground state of the system. The characteristic properties of the different phases can be calculated directly from this ground state wave 
function.

\subsection{Phase diagrams}
\subsubsection{$V_1/U=0, V_2/U\neq 0$}

\begin{figure}[tb]
\begin{center} 
\includegraphics[width=0.95\linewidth]{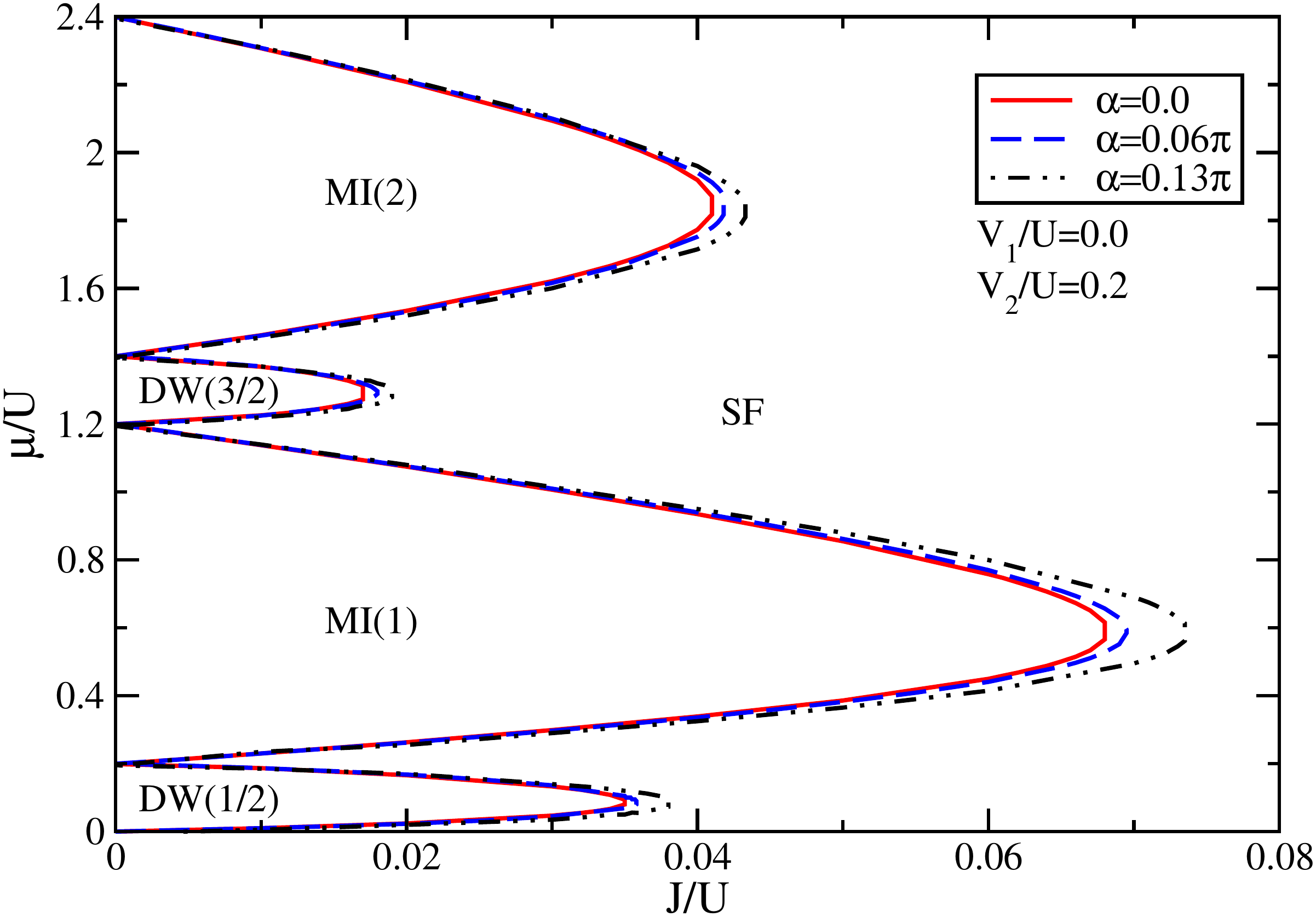} 
\end{center} 
\caption{(Color online) Phase diagram of the eBHM obtained from the CMFT method, for $V_1/U=0$, $V_2/U=0.2$ and $\alpha/\pi=0, 0.2$ and $0.4$.}
\label{fig:fig_man2}
\end{figure}

We first consider the case where the NNI along the legs of the ladder is zero, but finite along the rungs of the ladder. For this we fix 
$V_1/U=0$ and $V_2/U=0.2$ and calculate the ground state phase diagram for $\alpha=0, 0.2$ and $0.4$ (see Fig.~\ref{fig:fig_man2}). 
The repulsive NNI along the rungs of the ladder gives rise to DW and MI gapped phases and for sufficiently small values of J/U ($\lesssim 0.4$), as $\mu/U$ is increased, the system first enters the DW 
phase at density $\rho=0.5$, followed by MI phase at $\rho=1.0$. This pattern is then repeated with a DW phase at $\rho=1.5$ and a MI phase 
at $\rho=2.0$. As J/U is increased further, first the DW phases disappear and at still higher values of J/U the MI phases disappear as well.
The region in between the gapped phases and outside the lobes is occupied by the gapless superfluid phase.

\begin{figure}[tb]
\begin{center} 
\includegraphics[width=0.95\linewidth]{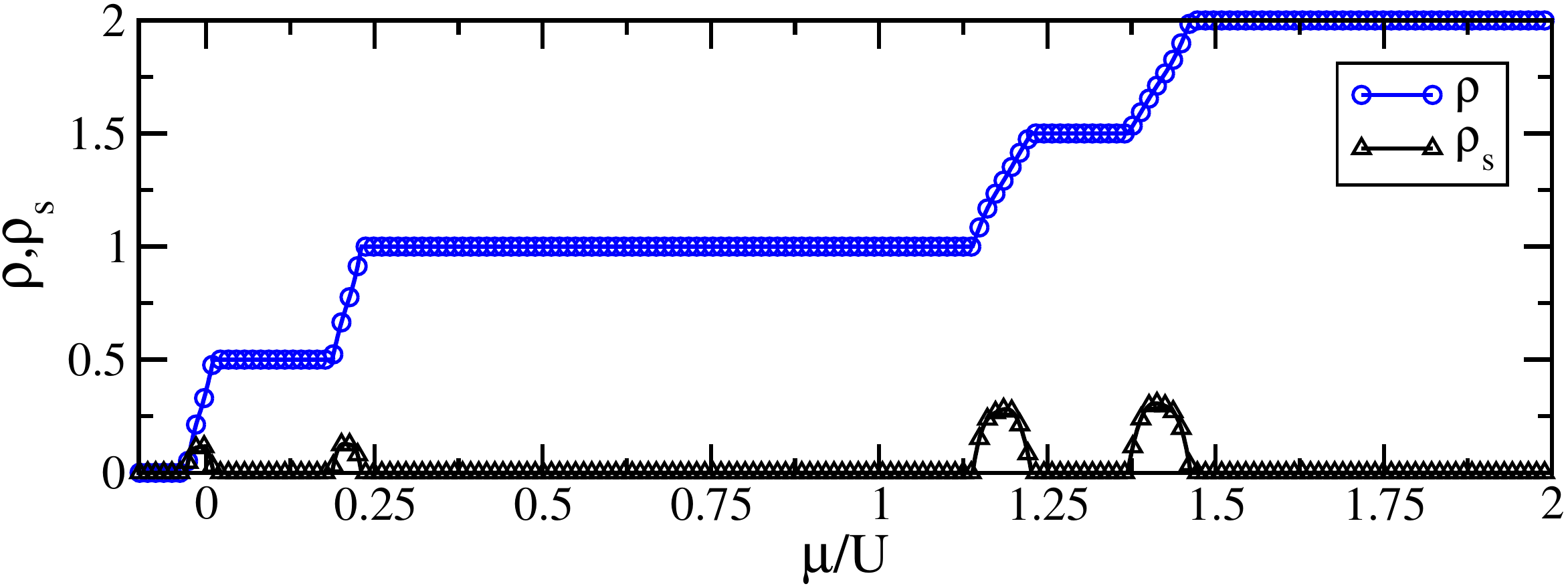} 
\end{center} 
\caption{(Color online)Densities $\rho$ and $\rho_s$ vs. $\mu/U$  for $J/U=K/U=0.1$, $V_1/U=0$, $V_2/U=0.2$ and $\alpha=0$.}
\label{fig:fig_man4-1}
\end{figure}

To better understand the origin of these two gapped phases we show the $\rho,\rho_s$ vs. $\mu$ phase diagram in
Fig.~\ref{fig:fig_man4-1}. Here $\rho$  indicates the average occupation number density and $\rho_s$ the superfluid density of the system.
When a finite and positive NNI is added along the rungs of the ladder, the particles in each rung start to repel 
each other and minimize the energy by redistributing themselves on different rungs.
Therefore, as the system approaches the density $\rho=0.5$, each rung 
has one particle in the minimum energy configuration and adding an extra particle to this configuration requires the energy $V_2$. This gives rise 
to the gapped phase, which is manifested in Fig.~\ref{fig:fig_man4-1} by the existence of a plateau at  $\rho=0.5$.
If the number of 
particles is commensurate with the number of rungs, they get distributed evenly. However, if this number is incommensurate, 
 extra particles exist which are free to hop and can be localized in any of the rungs, which in turn gives rise to the intermediate superfluid phase. 
Signatures of the superfluid phase can seen between the plateaus where the $\rho$ rises uniformly and where the superfluid 
density $\rho_s$ is finite.
As $\mu$ is increased further, the density of the system becomes one and it enters the MI phase where each site is occupied by one particle. 
At even higher $\mu$ the density increases to $1.5$ and each rung accommodates three particles. Each of these phases is gapped which is 
again confirmed by the presence of plateaus in $\rho$ and the absence of superfluid density. 
The additional effect of $\alpha$ on these phases is to expand the gapped regions, with larger $\alpha$ leading to larger regions. 
The results found from these calculations qualitatively match the results from the variational mean-field calculations above.

\begin{figure*}
\begin{center} 
\includegraphics[width=1.0\linewidth]{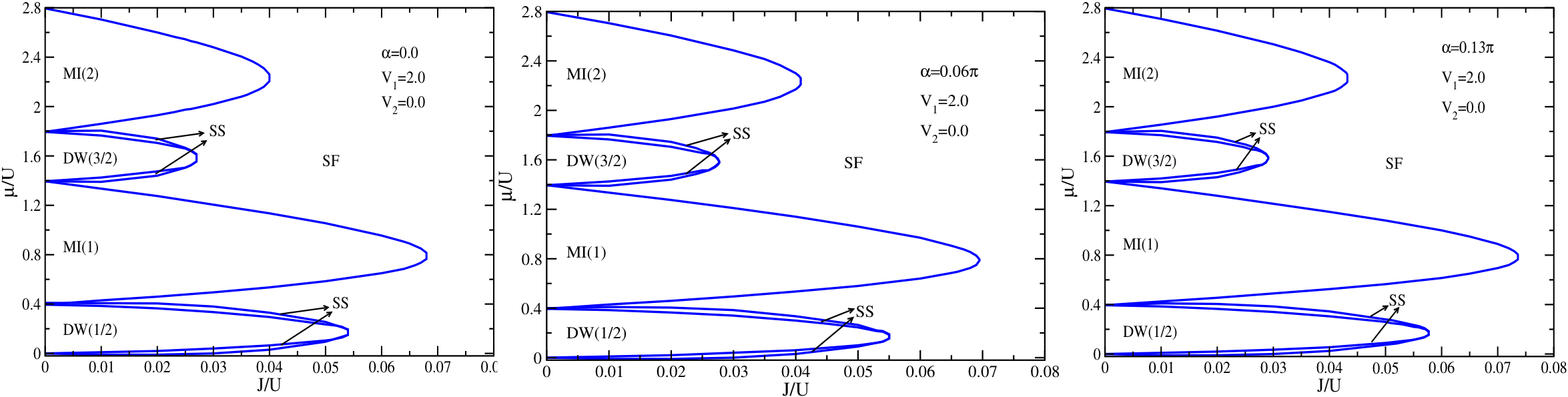} 
\caption{(Color online) $\mu/U$ vs. $J/U$ phase diagrams for $\alpha/\pi=0.0$ (left), $0.2$ (center) and $0.4$ (right) for
$V_1/U=0.2$ and $V_2/U=0$. One can see that the phases having densities of $1/2$ and $3/2$  are flanked by supersolid phases on both sides.}
\label{fig:3inone}
 \end{center} 
\end{figure*}

\subsubsection{$V_1/U\neq0, V_2/U= 0$}
\begin{figure}[tb]
\begin{center} 
\includegraphics[width=0.95\linewidth]{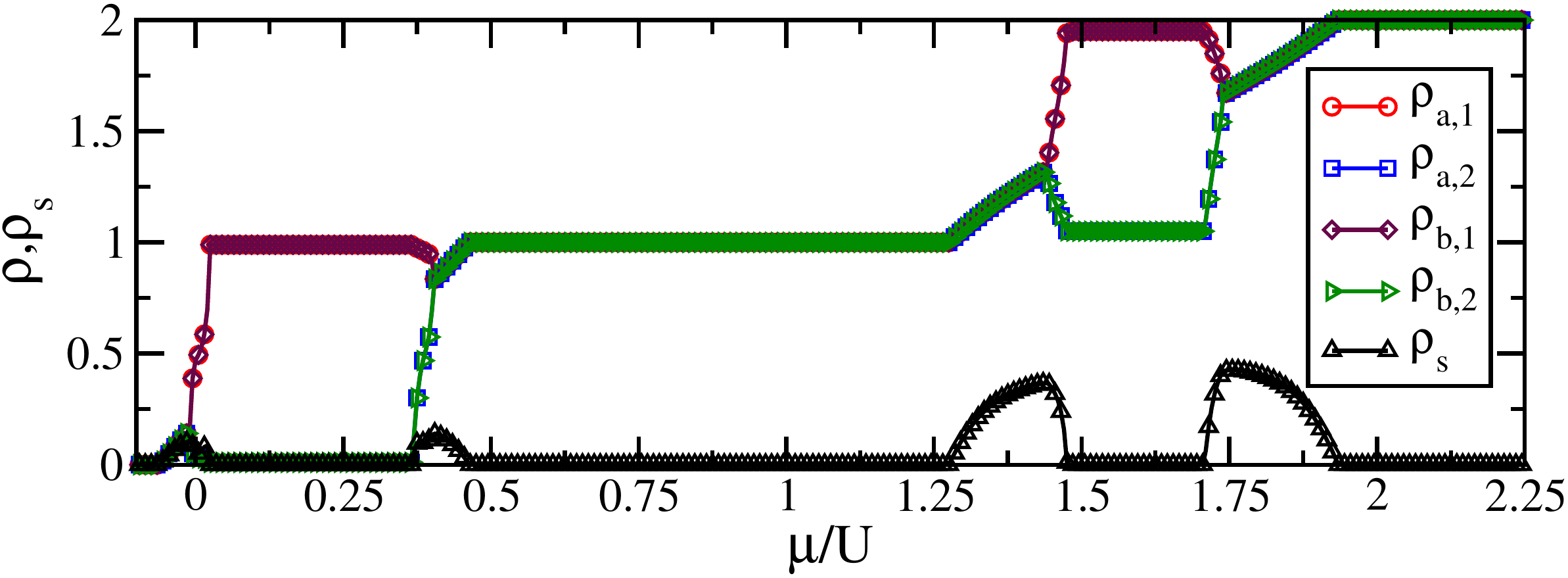} 
\end{center} 
\caption{(Color online) Densities $\rho$ and $\rho_s$ vs. $\mu/U$  for $J/U=K/U=0.2$, $V_1/U=0.2$, $V_2/U=0$, $\alpha=0$.}
\label{fig:fig_man4-2}
\end{figure}

After studying the effect of finite interaction $V_2$ along the rungs  on the phase diagram, we now focus on the NNI 
along the legs of the ladder, $V_1$.
To study this case we choose $V_1/U=0.2$ and $V_2/U=0$ and show the results in 
Figs.~\ref{fig:3inone} and \ref{fig:fig_man4-2}.  One can immediately see that,
compared to the previous case, a supersolid phase flanking the density-wave regions occurs. This can be understood by noting that in
the previous case the NNI was only present along the rungs of the ladder, which lead to a restricted hopping of the atoms in that direction.
However in the present case, the NNI is along the legs of the ladder, 
which means that no limitation on hopping along the rungs exist. This allows for off-diagonal long range order of the bosonic creation 
operator along with the density wave order, 
resulting in the formation of supersolid phase, see Fig.~\ref{fig:fig_man4-2}. In addition to the plateaus for gapped phases and the regular SF phase,
there are certain regions where SF density is finite and also there is a density imbalance between the rungs of the ladder. Red (circle), 
brown (diamond) density curves belong to the same rung (1) and blue (square), green (triangle) curves correspond to the density on the other 
rung (2). It can be seen that former as well as latter two curves overlap completely indicating that the sites on the same rung have same densities.

\subsubsection{$V_1/U\neq0, V_2/U\neq0$}
Finally, we examine the case when $J=K$, $V_1/U=V_2/U=0.2$ and $\alpha=0, 0.15\pi$ and $0.3\pi$, which can be compared to the analytical results shown in Fig.~\ref{fig_ebhmfull}.

\begin{figure}[tb]
\begin{center} 
\includegraphics[width=0.95\linewidth]{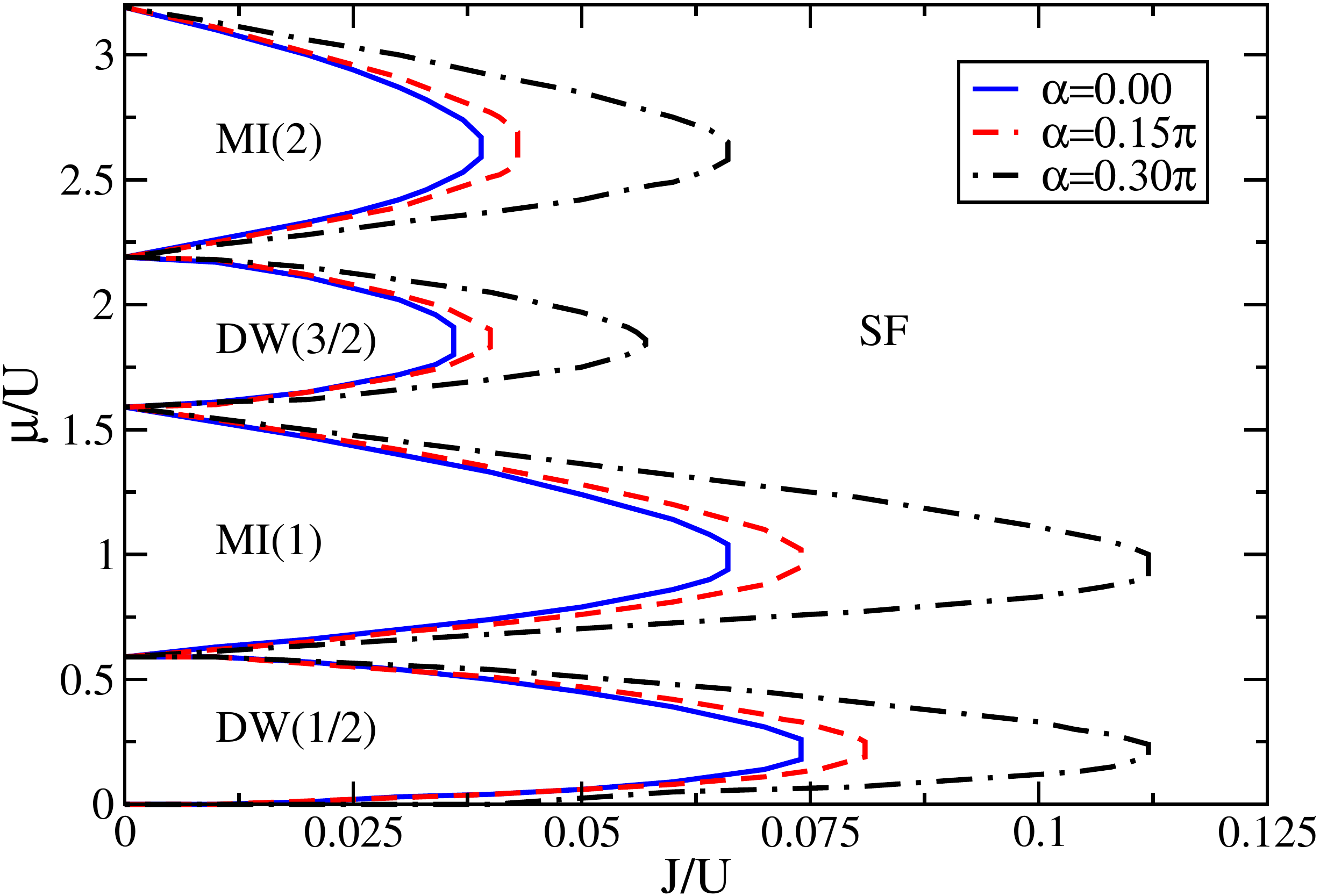} 
\end{center} 
\caption{(Color online) Phase diagram for different values of magnetic flux $\alpha$,
for $V_1/U=V_2/U=0.2$, calculated using the cluster mean field approach.}
\label{fig:fig_man7}
\end{figure}

Like before, we substitute these values in $H_c$ and solve it to find the ground state wavefunction. The phase diagram
obtained in this case is shown in Fig.~\ref{fig:fig_man7}, and it can be seen to be similar to the one Fig.~\ref{fig_ebhmfull}, obtained earlier. As $V_1$ and $V_2$ are both finite, the particles experience a repulsion along the legs as well as along the rungs of the ladder. Hence to minimize the energy the particles arrange themselves in way which is analogous to a checkerboard solid phase on a 
square lattice. In the DW phases, any two adjacent sites on the leg/rung of the ladder will have different number of particles
while the sites located diagonally in a plaquette will have same number of particles. In this case also the magnetic flux $\alpha$
enhances the gapped phases, and it confirms the results obtained using the mean field Gutzwiller approach. However as mentioned earlier, we cannot expect the results from Gutzwiller approach to compare with the cluster mean field theory. This is because unlike cluster mean field theory where correlations are included within each cluster, the analytical Gutzwiller approach does not include any correlations at all and hence, the location of transition boundaries differs in both cases. 

\section{Summary and outlook}\label{Sec: Summary}
We have examined the extended Bose Hubbard model on a two-legged ladder in the presence of a magnetic flux. We have shown that such a system possess a richly structured phase diagram, which is strongly influenced by the magnetic flux.  The presence of weak or strong nearest-neighbour interactions leads to the appearance of phases that are not present in the standard BH model, which include density-waves and supersolid states.  

We believe that this model serves as an important tool for understanding the fundamental properties of lattices gases coupled to gauge fields, and in particular helps elucidating the interplay between repulsive onsite and nearest neighbor interactions in lattice system. 

\appendix
\section{Minimization of energy functional for Mott-insulator/superfluid phase boundary} \label{misf_app}
The variational energy per rung of the system, $E=\langle\Psi|H|\Psi\rangle/\langle\Psi|\Psi\rangle$, up to second order in  $\Delta_{a}, \Delta_{a} ^{'},\Delta_{b}, \Delta_{b}^{'} $ can be calculated as
\begin{widetext}
\begin{align}
 E &= \{ -2J~\text{cos} (\alpha+\theta)[n_{0}\Delta_{a}^{2}+2\sqrt{n_0(n_0+1)} \Delta_{a}\Delta_{a}'
+(n_0+1)\Delta_{a}^{'2}]-2J~\text{cos} (\alpha-\theta)[\Delta_{b}^{2}n_{0}\nn\\
&  +2\sqrt{n_0(n_0+1)} \Delta_{b}\Delta_{b}'+(n_0+1)\Delta_{b}^{'2}]  -2K~[n_0\Delta_{a} \Delta_{b}+(n_0+1) \Delta_{a}' \Delta_{b}'\nn\\
&  +\sqrt{n_0(n_0+1)} \{\Delta_{a} \Delta_{b}'+\Delta_{a}'\Delta_{b}] +[(1-n_0)(\Delta_{a} ^{2}+\Delta_{b}^{2})+n_0(n_0+1)+n_0(\Delta_{a} ^{'2}+ \Delta_{b}^{'2})]\nn\\
&  +\mu [-2n_0+(\Delta_{a} ^{2}-\Delta_{a}^{'2})+(\Delta_{b} ^{2}-\Delta_{b}^{'2})]\},
\label{en_func_mi}
\end{align}
where the site indices are not written for compactness.
This energy has to be minimised with respect to the variational parameters $\Delta_{a}, \Delta_{a} ^{'},\Delta_{b}, \Delta_{b}^{'}$ and $\theta$ by calculating the Jacobian matrix of the second derivatives 
\[\mathbf{J}=-2J\left( \begin{array}{cc}
n_0\mathbf{F} & \sqrt{n_0(n_0+1)}\mathbf{F} \\
\sqrt{n_0(n_0+1)}\mathbf{F} & (n_0+1)\mathbf{F} \end{array} \right)
+2\left( \begin{array}{cc}
\{(1-n_0+\mu)-n_0(2V_1+V_2)\}\mathbf{I} & 0 \\
0 & \{(n_0-\mu)+n_0(2V_1+V_2)\}\mathbf{I} \end{array} \right),\]

where $\mathbf{I}$ is a $2\times2$ identity matrix and $ \mathbf{F}$ is given by
\[\mathbf{F}=\left( \begin{array}{cc}
2~\text{cos}(\alpha+\theta) & K/J \\
K/J & 2~\text{cos}(\alpha-\theta) \end{array} \right).\] 
This matrix possesses the same structure as that of a single particle Hamiltonian and for a minimum it has to be positive definite, which means that all eigenvalues have to be positive. 

To determine the eigenvalues, we adopt the following procedure~\cite{oktel_ladder}. Let $\lambda_{F}$ and $\vec{u}$ be the eigenvalues and eigenvectors of $\mathbf{F}$ i.e.~$ \mathbf{F}\vec{u}=\lambda_{F}\vec{u}$ and $\vec{v}=(a\vec{u}~~ b\vec{u})^{T}$ such that $ \mathbf{J}\vec{u}=\lambda\vec{u}$. Solving the eigenvalue equation, the roots can be found to be 

\bea \lambda_{1,2}&=&y_1\pm \sqrt{y_1^{2}-(n_0-\mu)(1-n_0+\mu)-J\lambda_F\{(1+\mu)-n_0(2V_1+V_2)\}+n_0(2V_1+V_2)\{(1-2n_0+2\mu)-n_0(2V_1+V_2)\}}\nn\\
\label{eigval}\eea

with $y_1=1-J(2n_0+1)\lambda_F$. Equating the minimum eigenvalue of the Jacobian matrix in Eq.~\eqref{eigval} to zero then gives the phase boundary of the Mott-insulator/superfluid phase. 

\section{Minimization of energy functional for Density-wave/supersolid phase boundary} \label{dwss_app}

The total variational energy up to second order in the variational parameters is given by
\begin{align} 
E &=  -2J~\text{cos} (\alpha+\theta)[\Delta_{a_1}\Delta_{a_2}\sqrt{n_{0}m_0}+\Delta_{a_1}\Delta_{a_2}'\sqrt{n_0(m_0+1)} +\Delta_{a_1}'\Delta_{a_2}'\sqrt{m_0(n_0+1)}+\Delta_{a_1}'\Delta_{a_2}'\sqrt{(m_0+1)(n_0+1)}]\nn\\
& -2J~\text{cos} (\alpha-\theta)[\Delta_{b_1}\Delta_{b_2}\sqrt{n_{0}m_0}+\Delta_{b_1}\Delta_{b_2}'\sqrt{n_0(m_0+1)} +\Delta_{b_1}'\Delta_{b_2}'\sqrt{m_0(n_0+1)}+\Delta_{b_1}'\Delta_{b_2}'\sqrt{(m_0+1)(n_0+1)}]\nn\\
&  -\frac{K}{2}[\Delta_{a_1}\Delta_{b_2}\sqrt{n_{0}m_0}+\Delta_{a_2}\Delta_{b_1}\sqrt{n_{0}m_0}+\Delta_{a_1}'\Delta_{b_2}'\sqrt{(n_0+1)(m_0+1)} +\Delta_{a_2}'\Delta_{b_1}'\sqrt{(n_0+1)(m_0+1)} \nn\\
&  +\Delta_{a_1}'\Delta_{b_2}\sqrt{m_0(n_0+1)}+\Delta_{a_1}\Delta_{b_2}'\sqrt{n_0(m_0+1)}+
\Delta_{a_2}'\Delta_{b_1}\sqrt{n_0(m_0+1)}+\Delta_{a_2}\Delta_{b_1}'\sqrt{m_0(n_0+1)}]\nn\\
&  +\frac{\mu}{2} [-2(n_0+m_0)+(\Delta_{a_1} ^{2}-\Delta_{a_1}^{'2})+(\Delta_{a_2} ^{2}-\Delta_{a_2}^{'2})+(\Delta_{b_1} ^{2}-\Delta_{b_1}^{'2})+(\Delta_{b_2} ^{2}-\Delta_{b_2}^{'2})]\}\nn\\
&  \frac{1}{2}[n_0(n_0-1)+m_0(m_0-1)+n_0(\Delta_{a_1} ^{'2}+\Delta_{b_1}^{'2})+m_0(\Delta_{a_2} ^{'2}+\Delta_{b_2}^{'2})+(1-n_0)(\Delta_{a_1} ^{2}+\Delta_{b_1}^{2})+(1-m_0)(\Delta_{a_2} ^{2}+\Delta_{b_2}^{2})] \nn\\
&  +\frac{(2V_1+V_2)}{2}[2 n_0m_0+n_0(\Delta_{a_2} ^{'2}-\Delta_{a_2}^{2}+\Delta_{b_2} ^{'2}-\Delta_{b_2}^{2})+m_0(\Delta_{a_1} ^{'2}-\Delta_{a_1}^{2}+\Delta_{b_1} ^{'2}-\Delta_{b_1}^{2})]\label{en_func_dw},
\end{align}
which needs to be minimized with respect to the eight variational parameters and $\theta$. We adopt the similar procedure as done for the Mott-insulator phase boundary, by calculating the Jacobian matrix of the second derivatives 
\[\mathbf{J}=\left( \begin{array}{cccc}
[\mu+1-n_0-m_0(2V_1+V_2)]\mathbf{I} & -2J\sqrt{n_0m_0}\mathbf{F} & 0 & -2J\sqrt{n_0(m_0+1)}\mathbf{F} \\
-2J\sqrt{n_0m_0}\mathbf{F} & [\mu+1-m_0-n_0(2V_1+V_2)]\mathbf{I} & -2J\sqrt{m_0(n_0+1)}\mathbf{F} & 0 \\
0 & -2J\sqrt{m_0(n_0+1)}\mathbf{F}  & [-\mu+n_0+m_0(2V_1+V_2)]\mathbf{I} &  -2J\sqrt{(n_0+1)(m_0+1)}\mathbf{F} \\
 -2J\sqrt{n_0(m_0+1)}\mathbf{F} & 0 &  -2J\sqrt{(n_0+1)(m_0+1)}\mathbf{F} &   [-\mu+m_0+n_0(2V_1+V_2)]\mathbf{I}
 \end{array} \right)
  \] 

where $\mathbf{I}$ is a $2\times2$ identity matrix and $ \mathbf{F}$ is as above. The eigenvalue equation then takes the form
\bea (\lambda-2T_1)(\lambda-2T_2)(\lambda-2T_3)(\lambda-2T_4)=4J^2\lambda_F^2[n_0(\lambda-2T_2)+(n_0+1)(\lambda-2T_1)]~[m_0(\lambda-2T_4)+(m_0+1)(\lambda-2T_3)], \nn\\
\label{eig_eq_dw}\eea
\end{widetext}
which is a fourth order equation in $\lambda$. The next step is to determine the roots of this equation and we make use of the smallest root by setting it equal to zero, which gives  the value of the critical hopping amplitude $J_c$ for any value of $\mu$ and magnetic flux $\alpha$ in the ladder. The results for the critical hopping amplitudes are plotted in Figs.~\ref{fig:V_R}, and \ref{fig_ebhmfull} as a function of chemical potential $\mu$ and the magnetic flux $\alpha$ through the ladder.

\section*{Acknowledgements}  This work was supported by the Okinawa Institute of Science and Technology Graduate University, Okinawa, Japan. M.S. would
like to thank Physical Research Laboratory, Ahmedabad for providing the HPC facilities on Vikram-100. 
% M.S. would like to acknowledge DST-SERB, India for the financial support through Project No. PDF/2016/000569. 

\end{document}